\documentclass{article}
\usepackage{spconf,amsmath,graphicx}
\usepackage{amssymb,times,url,bm}
\usepackage{color}
\usepackage[printonlyused, nolist]{acronym}
\usepackage{algorithm,algorithmic}
\usepackage{tikz,pgfplots}
\usepackage{booktabs}

\newcommand{\MixingMat}{\mathbf{A}}
\newcommand{\SteeringVector}{\mathbf{h}}

\newcommand{\W}{\mathbf{W}}
\newcommand{\w}{\mathbf{w}}
\newcommand{\invW}{\mathbf{W}^{-1}}
\newcommand{\y}{\mathbf{y}}
\newcommand{\x}{\mathbf{x}}
\newcommand{\s}{\mathbf{s}}
\newcommand{\V}{\mathbf{V}}

\newcommand{\FreqIdx}{f}
\newcommand{\FreqIdxMax}{F}
\newcommand{\BlockIdx}{n}
\newcommand{\BlockIdxMax}{N}
\newcommand{\ChannelIdx}{k}
\newcommand{\ChannelIdxMax}{K}
\newcommand{\yFreqVec}{\underline{\y}}

\newcommand{\SetW}{\mathcal{W}}

\newcommand{\Y}{\mathcal{Y}}
\newcommand{\X}{\mathcal{X}}

\newcommand{\SetN}{\mathcal{N}}
\newcommand{\SetK}{\mathcal{F}}
\newcommand{\SetP}{\mathcal{K}}

\newcommand{\transp}{^\text{T}}
\newcommand{\herm}{^\text{H}}
\newcommand{\Expect}[1]{\hat{\mathbb{E}}\left\lbrace #1 \right\rbrace}

\newlength\fwidth

\newcommand{\Walter}[1]{#1}
\newcommand{\WalterTwo}[1]{#1}

\newcommand{\WASPAA}[1]{#1}

\title{Spatially Guided Independent Vector Analysis}

\name{Andreas Brendel, Thomas Haubner and Walter Kellermann\thanks{This work was supported by DFG under contract no $<$Ke890/10-1$>$ within the Research Unit FOR2457 "Acoustic Sensor Networks"}}
\address{Multimedia Communications and Signal Processing, Friedrich-Alexander-Universit\"at Erlangen-N\"urnberg,\\
	Cauerstr. 7, D-91058 Erlangen, Germany, \texttt{Andreas.Brendel@FAU.de}}
\begin{document}
		\begin{acronym}
			\acro{STFT}{Short-Time Fourier Transform}
			\acro{PSD}{Power Spectral Density}
			\acro{PDF}{Probability Density Function}
			\acro{RIR}{Room Impulse Response}
			\acro{FIR}{Finite Impulse Response}
			\acro{FFT}{Fast Fourier Transform}
			\acro{DFT}{Discrete Fourier Transform}
			\acro{ICA}{Independent Component Analysis}
			\acro{IVA}{Independent Vector Analysis}
			\acro{TRINICON}{TRIple-N Independent component analysis for CONvolutive mixtures}
			\acro{FD-ICA}{Frequency Domain ICA}
			\acro{BSS}{Blind Source Separation}
			\acro{NMF}{Nonnegative Matrix Factorization}
			\acro{MM}{Majorize-Minimize}
			\acro{MAP}{Maximum A Posteriori}
			\acro{RTF}{Relative Transfer Function}
			\acro{auxIVA}{Auxiliary Function IVA}
			\acro{FD-ICA}{Frequency-Domain Independent Component Analysis}
			\acro{DOA}{Direction of Arrival}
			\acro{SNR}{Signal-to-Noise Ratio}
			\acro{SIR}{Signal-to-Interference Ratio}
			\acro{SDR}{Signal-to-Distortion Ratio}
			\acro{GC}{Geometric Constraint}
			\acro{DRR}{Direct-to-Reverberant energy Ratio}
			\acro{ILRMA}{Independent Low-Rank Matrix Analysis}
		\end{acronym}
\ninept
\maketitle
		\begin{abstract}
			\WASPAA{We present a Maximum A Posteriori (MAP) derivation of the Independent Vector Analysis (IVA) algorithm for blind source separation incorporating an additional spatial prior over} the demixing matrices. In this way, the outer permutation ambiguity of IVA is avoided \WASPAA{and the algorithm can be guided towards a desired solution in adverse acoustic conditions}. The resulting MAP optimization problem is solved by deriving majorize-minimize update rules to achieve convergence speed comparable to the well-known auxiliary function IVA algorithm\WASPAA{, i.e., the convergence is not impaired by the additional constraint}. \WASPAA{The proposed algorithm exhibits superior performance at lower computational cost than a state-of-the-art spatially constrained IVA algorithm in a setup defined by} \WASPAA{real-world Room Impulse Responses (RIRs)}.
		\end{abstract}
		
		\begin{keywords}
			Independent Vector Analysis, MM Algorithm, Directional Constraint
		\end{keywords}
		
		\section{Introduction}
		\label{sec:intro}
		\ac{BSS}, i.e., the estimation of signals out of a recorded mixture with only little information about the underlying scenario, is a core task of audio signal processing problems and has been addressed in a multitude of proposed approaches in the last decades \WASPAA{\cite{vincent_audio_2018,pedersen_survey_2007}}. For the most practically relevant scenario of a convolutive mixture, \ac{FD-ICA} \cite{smaragdis_blind_1998} has been proposed which estimates demixing matrices independently in each frequency \WASPAA{bin} such that the output signals are statistically independent. However, this causes the well-known inner permutation problem \cite{sawada_robust_2004}, which has to be resolved afterwards. As a method which avoids the inner permutation problem by choosing a multivariate source prior over all frequency \WASPAA{bins}, \ac{IVA} \cite{kim_blind_2007} has attracted much attention. Based on the \ac{MM} principle \cite{hunter_tutorial_2004}, stable and fast update rules, named \ac{auxIVA} \cite{vigneron_auxiliary-function-based_2010,ono_stable_2011}, have been derived which do not require any tuning parameter, \Walter{e.g.}, a step-size. \WASPAA{Various ways to incorporate prior knowledge into \ac{IVA} have been proposed, whereby knowledge about the source variances is the most established one \cite{ono_user-guided_2012}. \ac{NMF} \cite{lee_algorithms_2000} is used in the \ac{ILRMA} algorithm \cite{kitamura_determined_2016} to estimate these source variances.}
		
		Another ambiguity, which is inherent to \ac{BSS}, is the ordering of the \Walter{broadband} signals at the output channels, i.e., the outer permutation ambiguity. Prior knowledge has to be introduced to solve this issue by guiding the adaptation of the demixing filters. To this end, supervised \ac{IVA} \cite{nesta_supervised_2017} has been proposed, which introduces pilot signals \Walter{that} are statistically dependent on the source signals into a gradient-based update rule. \Walter{Another} idea, which has been successfully applied for resolving the outer permutation problem, is to exploit spatial information \Walter{about} the sources. \Walter{Such techniques include exploiting} the dominance of a source for a certain direction \Walter{in \ac{FD-ICA}} \cite{sawada_blind_2005}, initialization \Walter{of \ac{auxIVA}} with filters obeying a free-field model \cite{chen_auxiliary_2014}, \Walter{using} prelearned filters \Walter{for gradient-based \ac{IVA}} \cite{koldovsky_semi-blind_2013} or imposing a \ac{GC} \cite{parra_geometric_2002,knaak_geometrically_2007,zhang_combining_2009,reindl_minimum_2014,makino_informed_2018} \Walter{on different \ac{BSS} variants}. For \ac{IVA}, a \WalterTwo{geometrically-constrained} \Walter{gradient-based} update rule has been proposed in \cite{vincent_geometrically_2015}.
		
		In this contribution, we provide a \ac{MAP} derivation of \ac{IVA}, based on the previous work for \ac{ICA} \cite{knuth_bayesian_1999}, which allows to incorporate prior knowledge about the demixing system via a prior \ac{PDF} \WASPAA{to preclude the outer permutation problem and guide the algorithm to a desired solution in acoustically demanding scenarios}. This allows to express the uncertainty of the localization information and to fuse the proposed \ac{MAP} \ac{IVA} with a localization or tracking algorithm by exploiting the uncertainties of the estimates. Finally, motivated by the tremendous \Walter{advantage regarding} convergence speed of \ac{auxIVA} \cite{ono_stable_2011} in comparison to gradient-based \ac{IVA} \cite{kim_blind_2007}, we derive update rules based on the \ac{MM} principle providing faster convergence than the competing gradient-based methods without the necessity for tuning the step size \WASPAA{or impairing the convergence relative to \ac{IVA}}. \WASPAA{Note that we do not compare our approach with \ac{ILRMA} as for \ac{ILRMA} only prior spectral knowledge about the sources can be introduced if used in a semi-supervised setup, but no spatial prior knowledge, which is the focus of this paper.}
		
		In the following, scalar variables are denoted by lower-case letters, vectors by bold lower-case letters, matrices by bold upper-case letters and sets as calligraphic upper-case letters. $[\cdot]_i$ or $[\cdot]_{i,j}$ denotes the $i$th element of a vector or the element in the $i$th row and $j$th column of a matrix\Walter{,} and $(\cdot)\transp$, \WASPAA{$(\cdot)^\ast$} and $(\cdot)\herm$ denote transposition, \WASPAA{complex conjugation} and hermitian, respectively.
		
		\section{Probabilistic Model}
		\label{sec:prob_model}
		In the following, we \Walter{study} a determined scenario, i.e., the number of sources equals the number of sensors $\ChannelIdxMax$. \Walter{Assuming} sufficiently shorter impulse responses between sources and microphones than the window length of the \ac{STFT}, the microphone signals can be described at time step \mbox{$\BlockIdx \in \SetN = \{1,\dots,\BlockIdxMax\}$}, \Walter{where $\BlockIdxMax$ is the number of \WalterTwo{observed} time frames}, and frequency index $\FreqIdx \in \SetK = \{1,\dots,\FreqIdxMax\}$ as \vspace{-2pt}
		\begin{equation}
			\x_{\FreqIdx,\BlockIdx} = \MixingMat_\FreqIdx \s_{\FreqIdx,\BlockIdx}. \vspace{-2pt}
		\end{equation}
		Hereby, $\MixingMat_\FreqIdx \in \mathbb{C}^{\ChannelIdxMax\times\ChannelIdxMax}$ is the matrix of acoustic transfer functions at frequency index $\FreqIdx$ and \vspace{-2pt}
		\begin{equation}
			\s_{\FreqIdx,\BlockIdx} = \left[s_{\FreqIdx,\BlockIdx}^1,\dots,s_{\FreqIdx,\BlockIdx}^\ChannelIdxMax\right]\transp, \x_{\FreqIdx,\BlockIdx} = \left[x_{\FreqIdx,\BlockIdx}^1,\dots,x_{\FreqIdx,\BlockIdx}^\ChannelIdxMax\right]\transp \in \mathbb{C}^{\ChannelIdxMax}
		\end{equation}
		denote the input signals and microphone signals with channel or signal index $\ChannelIdx \in \SetP = \{1,\dots,\ChannelIdxMax\}$, respectively. An estimate of the demixed signals \vspace{-7pt}
		\begin{equation}
			\y_{\FreqIdx,\BlockIdx} = \left[y_{\FreqIdx,\BlockIdx}^1,\dots,y_{\FreqIdx,\BlockIdx}^\ChannelIdxMax\right]\transp  \in \mathbb{C}^{\ChannelIdxMax}
		\end{equation}
		can be obtained by applying a demixing matrix \WASPAA{for each} frequency \mbox{\WASPAA{bin} $\FreqIdx$} \vspace{-2pt}
		\begin{equation}
			\W_\FreqIdx = \begin{bmatrix}
				\w_\FreqIdx^1,\dots,\w_\FreqIdx^\ChannelIdxMax
			\end{bmatrix}\herm \in \mathbb{C}^{\ChannelIdxMax\times\ChannelIdxMax},
		\end{equation}
		\Walter{\WalterTwo{representing the $\ChannelIdxMax$} demixing filters for \WalterTwo{the $\ChannelIdx$th output in vector} \WASPAA{$(\w_\FreqIdx^\ChannelIdx)\herm$},} to the observed microphone signals \vspace{-2pt}
		\begin{equation}
			\y_{\FreqIdx,\BlockIdx} = \W_\FreqIdx \x_{\FreqIdx,\BlockIdx}.
			\label{eq:demixing_equation}
		\end{equation}
		Additionally, we define the demixed \Walter{broadband} signal vector for channel $\ChannelIdx$ over all frequencies and \Walter{the} concatenation as \vspace{-2pt}
		\begin{equation}
			\yFreqVec_{\ChannelIdx,\BlockIdx} = \left[y_{1,\BlockIdx}^\ChannelIdx,\dots,y_{\FreqIdxMax,\BlockIdx}^\ChannelIdx\right]\transp  \in \mathbb{C}^{\FreqIdxMax}, \yFreqVec_{\BlockIdx} = \left[\yFreqVec_{1,\BlockIdx}\transp,\dots,\yFreqVec_{\ChannelIdxMax,\BlockIdx}\transp\right]\transp \in \mathbb{C}^{\ChannelIdxMax\FreqIdxMax},
		\end{equation}
		respectively. The set of all demixing matrices is denoted as \mbox{$\SetW = \left\lbrace\W_\FreqIdx \in \mathbb{C}^{\ChannelIdxMax\times\ChannelIdxMax}\vert \FreqIdx \in \SetK\right\rbrace$}, the set of all demixed signal vectors as $\Y = \left\lbrace\y_\BlockIdx \in \mathbb{C}^{\ChannelIdxMax \FreqIdxMax}\vert \BlockIdx \in \SetN\right\rbrace$ and the set of all microphone observations as $\X = \left\lbrace\x_{\FreqIdx,\BlockIdx} \in \mathbb{C}^{\ChannelIdxMax}\vert \FreqIdx \in \SetK,\BlockIdx \in \SetN\right\rbrace$.
		
		Equipped with these definitions, we apply Bayes theorem to calculate the joint posterior of \WASPAA{the} demixed \Walter{broadband} signals and demixing matrices \vspace{-2pt}
		\begin{align}
			p(\SetW,\Y\vert \X) &= p(\SetW,\Y) \frac{p(\X|\SetW,\Y)}{p(\X)}= p(\SetW)p(\Y\vert\SetW) \frac{p(\X|\SetW,\Y)}{p(\X)} \notag\\
			&\propto p(\SetW)p(\Y\vert\SetW) p(\X|\SetW,\Y).
		\end{align}
		According to the deterministic relationship between microphone signals and demixed signals \eqref{eq:demixing_equation}, we model the likelihood for one \Walter{observed} time frame \WASPAA{index} $\BlockIdx$ and frequency \WASPAA{bin} $\FreqIdx$ to be \vspace{-2pt}
		\begin{equation}
			p\left(\x_{\FreqIdx,\BlockIdx} \big\vert \SetW,\y_{\FreqIdx,\BlockIdx}\right) = \delta \left(\x_{\FreqIdx,\BlockIdx}-\invW_\FreqIdx \y_{\FreqIdx,\BlockIdx}\right),
		\end{equation}
		assuming that the inverse \Walter{of $\W_\FreqIdx$} exists. Hereby, $\delta(\cdot)$ denotes the Dirac \Walter{distribution}. Furthermore, we assume independence between \WASPAA{time} blocks and frequency \WASPAA{bins}, which yields the likelihood \vspace{-2pt}
		\begin{align}
			p(\X|\SetW,\Y) &= \prod_{\BlockIdx=1}^{\BlockIdxMax}\prod_{\FreqIdx=1}^{\FreqIdxMax} \delta \left(\x_{\FreqIdx,\BlockIdx}-\invW_\FreqIdx \y_{\FreqIdx,\BlockIdx}\right).
		\end{align}
		The \ac{PDF} of all demixed signal vectors is obtained by assuming independence over all time blocks $\BlockIdx$ and signals $\ChannelIdx$ \vspace{-2pt}
		\begin{equation}
			p(\Y\vert\SetW) = \prod_{\BlockIdx=1}^{\BlockIdxMax}p\left(\yFreqVec_{\BlockIdx}\right) = \prod_{\BlockIdx=1}^{\BlockIdxMax}\prod_{\ChannelIdx=1}^{\ChannelIdxMax}p\left(\yFreqVec_{\ChannelIdx,\BlockIdx}\right).
		\end{equation}
		Note that $p(\yFreqVec_{\ChannelIdx,\BlockIdx})$ is a multivariate density \Walter{capturing} all \Walter{frequency bins}. Now, we compute the posterior of the demixing matrices by marginalizing the demixed signals \vspace{-2pt}
		\begin{align}
			p(\SetW|\X) &= \int p(\SetW, \Y|\X) d\yFreqVec_1\dots d\yFreqVec_\BlockIdxMax \notag\\
			&\propto p(\SetW) \int p(\Y\vert\SetW) p(\X|\SetW,\Y) d\yFreqVec_1\dots d\yFreqVec_\BlockIdxMax \notag\\
			&= p(\SetW) \prod_{\BlockIdx=1}^{\BlockIdxMax} \int p(\yFreqVec_{\BlockIdx}) \prod_{\FreqIdx=1}^{\FreqIdxMax} \delta \left(\x_{\FreqIdx,\BlockIdx}-\invW_\FreqIdx \y_{\FreqIdx,\BlockIdx}\right) d\yFreqVec_\BlockIdx \notag\\
			&= p(\SetW) \prod_{\FreqIdx=1}^{\FreqIdxMax} \vert \det \W_\FreqIdx \vert^{\WASPAA{2}\BlockIdxMax}\prod_{\BlockIdx=1}^{\BlockIdxMax} \prod_{\ChannelIdx=1}^{\ChannelIdxMax}p\left(\yFreqVec_{\ChannelIdx,\BlockIdx}\right),
		\end{align}
		\WASPAA{where we} used the sifting property of the Dirac \Walter{distribution} in the last step. Finally, we obtain the following \ac{MAP} optimization problem for the estimation of the demixing matrices \vspace{-7pt}
		\begin{align}
			\W_\FreqIdx &= \underset{\W_\FreqIdx \in \mathbb{C}^{\ChannelIdxMax\times\ChannelIdxMax}}{\arg\max}\,\frac{\log p(\SetW)}{\BlockIdxMax} + \WASPAA{2}\sum_{\FreqIdx=1}^{\FreqIdxMax} \log \vert \det \W_\FreqIdx \vert \dots \notag\\
			& \qquad \dots - \sum_{\ChannelIdx=1}^{\ChannelIdxMax} \Expect{G\left(\yFreqVec_{\ChannelIdx,\BlockIdx}\right)}. \vspace{-10pt}
			\label{eq:MAP_problem}
		\end{align}
		Here, we introduced the source model $G(\yFreqVec_{\ChannelIdx,\BlockIdx}) = -\log p(\yFreqVec_{\ChannelIdx,\BlockIdx})$ and the averaging operator
		$\Expect{\cdot} = \frac{1}{\BlockIdxMax} \sum_{\BlockIdx=1}^{\BlockIdxMax}(\cdot)$.\vspace{-3pt}
		\subsection{Relation to IVA}
		\label{sec:relation_IVA}
		By choosing \WASPAA{an} \Walter{uninformative} prior for the demixing matrices, $p(\SetW) = \text{const.}$,	and negating the maximization problem \eqref{eq:MAP_problem}, we arrive at the \ac{IVA} cost function \cite{kim_blind_2007,kitamura_effective_2017} \vspace{-2pt}
		\begin{equation}
			J_\text{IVA}(\SetW) = \sum_{\ChannelIdx=1}^{\ChannelIdxMax} \Expect{G\left(\yFreqVec_\ChannelIdx\right)} - \WASPAA{2}\sum_{\FreqIdx=1}^{\FreqIdxMax}\log \left\vert \det \W_\FreqIdx\right\vert, \vspace{-4pt}
			\label{eq:IVA_cost_function}
		\end{equation}
		i.e., the \ac{MAP} optimization problem yields the original \ac{IVA} cost function as a special case. \vspace{-3pt}
		\subsection{Choice of Prior \ac{PDF}}
		\label{sec:choice_prior}
		Assuming free-field propagation, the \WASPAA{$m$th} element of the \ac{RTF} $\SteeringVector_\FreqIdx^{\WASPAA{\ChannelIdx}}\WASPAA{\in \mathbb{C}^\ChannelIdxMax}$ in frequency \WASPAA{bin} $\FreqIdx$ w.r.t. the first microphone is \Walter{expressed} as  \vspace{-2pt}
		\begin{equation}
			[\SteeringVector_\FreqIdx^{\WASPAA{\ChannelIdx}}]_{\WASPAA{m}} = \left[\exp\left(j\frac{2\pi \nu_\FreqIdx}{c_s}\Vert\mathbf{r}_{\WASPAA{m}}-\mathbf{r}_1\Vert_2\cos \vartheta_{\WASPAA{\ChannelIdx}}\right)\right]_{\WASPAA{m}}. \vspace{-2pt}
		\end{equation}
		Hereby, $\nu_\FreqIdx$ denotes the frequency in $\mathrm{Hz}$ corresponding to frequency bin $\FreqIdx$, $c_s$ the speed of sound, $\mathbf{r}_{\WASPAA{m}}$ the position of the $\WASPAA{m}$th microphone, $\vartheta_\ChannelIdx$ the \ac{DOA} of the \Walter{considered} source \WASPAA{at channel $\ChannelIdx$} and $\Vert \cdot \Vert_2$ the Euclidean norm.
		We \WASPAA{model} the prior over the demixing matrices to be i.i.d. over all frequency \WASPAA{bins} \WASPAA{and channels} \vspace{-2pt}
		\begin{equation}
			p(\SetW) = \prod_{\FreqIdx=1}^{\FreqIdxMax}p(\W_\FreqIdx)\WASPAA{= \prod_{\FreqIdx=1}^{\FreqIdxMax}\prod_{\ChannelIdx=1}^{\ChannelIdxMax}p\left(\w_\FreqIdx^\ChannelIdx\right)}. \vspace{-2pt}
		\end{equation}
		\WASPAA{We propose the following prior, which favors a spatial null into the specified \ac{DOA} $\vartheta_\ChannelIdx$} \vspace{-2pt}
		\WASPAA{\begin{equation}
			p\left(\w_\FreqIdx^\ChannelIdx\right) = \frac{\exp\left(-\frac{1}{\tilde{\sigma}^2_\FreqIdx}(\w_\FreqIdx^\ChannelIdx)\herm \left(\lambda_\text{E}\mathbf{I}+\SteeringVector_\FreqIdx^{\WASPAA{\ChannelIdx}}(\SteeringVector_\FreqIdx^{\WASPAA{\ChannelIdx}})\herm\right)\w_\FreqIdx^\ChannelIdx\right)}{\sqrt{\left(\pi \tilde{\sigma}^2_\FreqIdx\right)^\ChannelIdxMax \det(\lambda_\text{E}\mathbf{I}+\SteeringVector_\FreqIdx^{\WASPAA{\ChannelIdx}}(\SteeringVector_\FreqIdx^{\WASPAA{\ChannelIdx}})\herm)}}. \vspace{-4pt}
			\label{eq:directional_prior}
		\end{equation}}
		\WASPAA{Here, the variable} $\tilde{\sigma}_\FreqIdx^2$ is a user-defined parameter, expressing the uncertainty of the \ac{DOA} estimate. However, $\tilde{\sigma}_\FreqIdx^2$ could be directly obtained from a localization or tracking algorithm. \WASPAA{The identity matrix in \eqref{eq:directional_prior} acts as a Tikhonov regularizer \cite{bishop_pattern_2006}, controlled by the parameter $\lambda_\text{E}$, i.e., this term is penalizing the filters energy.}
		\WASPAA{In the following we constrain the channels corresponding to the indices in the set $\mathcal{I}$ and choose a non-informative prior otherwise.}
		\section{Derivation of Update Rules}
		\label{sec:update}
		The cost function corresponding to the \ac{MAP} problem \eqref{eq:MAP_problem} and the chosen prior \ac{PDF} in Sec.~\ref{sec:choice_prior} is obtained as \vspace{-2pt}
		\begin{align}
			J(\SetW) &= \underbrace{\sum_{\ChannelIdx=1}^{\ChannelIdxMax}\hat{\mathbb{E}}\left\lbrace G\left(\yFreqVec_\ChannelIdx\right) \right\rbrace - \WASPAA{2}\sum_{\FreqIdx=1}^{\FreqIdxMax}\log \left\vert \det \W_\FreqIdx\right\vert}_{J_\text{IVA}(\SetW)}+\dots \notag\\
			&\qquad \dots \WASPAA{+\underbrace{\frac{1}{\sigma^2_\FreqIdx}\sum_{\FreqIdx=1}^{\FreqIdxMax}\sum_{\ChannelIdx=1}^{\ChannelIdxMax}(\w_\FreqIdx^\ChannelIdx)\herm \left(\lambda_\text{E}\mathbf{I}+\SteeringVector_\FreqIdx^{\WASPAA{\ChannelIdx}}(\SteeringVector_\FreqIdx^{\WASPAA{\ChannelIdx}})\herm\right)\w_\FreqIdx^\ChannelIdx}_{J_\text{prior}(\SetW)}}, \vspace{-2pt}
			\label{eq:proposed_cost_function}
		\end{align}
		where $\sigma^2_\FreqIdx = \BlockIdxMax \tilde{\sigma}_\FreqIdx^2$. The cost function \eqref{eq:proposed_cost_function} is composed of the summation of the original \ac{IVA} cost function $J_\text{IVA}(\SetW)$ and a nonnegative term corresponding to the contribution of the prior $J_\text{prior}(\SetW)\geq0$. In the following, we will derive an \ac{MM} algorithm \cite{hunter_tutorial_2004} based on \cite{ono_stable_2011} to minimize the proposed cost function \eqref{eq:proposed_cost_function}.
		
		\subsection{Construction of an Upper Bound}
		In the following, $\SetW^{(l)}$ \Walter{marks the set of estimated demixing matrices} at iteration $l\in\{1,\dots,L\}$ with \Walter{$L$ as the} maximum number of iterations. Furthermore, $Q(\SetW\vert\SetW^{(l)})$ denotes an upper bound of the cost function $J(\SetW)$ \Walter{at the $l$th iteration}. To develop an \ac{MM} algorithm for the optimization of the demixing matrices $\SetW$, we have to construct $Q(\SetW\vert\SetW^{(l)})$ such that it dominates the cost function for all choices of $\SetW$ \vspace{-2pt}
		\begin{equation}
			J(\SetW) \leq Q(\SetW\vert\SetW^{(l)})
		\end{equation}
		and is identical to the cost function iff $\SetW = \SetW^{(l)}$, i.e., \vspace{-2pt}
		\begin{equation}
			J(\SetW^{(l)}) = Q(\SetW^{(l)}\vert\SetW^{(l)}).
		\end{equation}
		Defining $\SetW_\ChannelIdx = \left\lbrace\w_\FreqIdx^\ChannelIdx \in \mathbb{C}^{\ChannelIdxMax}\vert \FreqIdx \in \SetK\right\rbrace$ as the set of all demixing vectors for source $\ChannelIdx$, we can \Walter{use} the following inequality for super-Gaussian source models $G$, which has been \Walter{proven} in \cite{vigneron_auxiliary-function-based_2010,ono_stable_2011} \vspace{-2pt}
		\begin{equation}
			\Expect{G\left(\yFreqVec_{\ChannelIdx,\BlockIdx}\right)} \leq \frac{1}{2}\sum_{\FreqIdx=1}^{\FreqIdxMax}\left(\left(\w_\FreqIdx^\ChannelIdx\right)\herm \V_\FreqIdx^\ChannelIdx\left(\SetW_\ChannelIdx^{(l)}\right)\w_\FreqIdx^\ChannelIdx\right) + R_\ChannelIdx\left(\SetW_\ChannelIdx^{(l)}\right).
			\label{eq:ono_inequality}
		\end{equation}
		\WalterTwo{Hereby, $R_\ChannelIdx(\SetW_\ChannelIdx^{(l)})$ constitutes a term which is independent of $\SetW$ \cite{vigneron_auxiliary-function-based_2010,ono_stable_2011}, and $\V_\FreqIdx^\ChannelIdx$ denotes} the weighted microphone signal covariance matrix \vspace{-2pt}
		\begin{equation}
			\V_\FreqIdx^\ChannelIdx\left(\SetW_\ChannelIdx^{(l)}\right) = \Expect{\frac{G'(r_{\BlockIdx}^\ChannelIdx(\SetW_\ChannelIdx^{(l)}))}{r_{\BlockIdx}^\ChannelIdx(\SetW_\ChannelIdx^{(l)})}\x_{\FreqIdx,\BlockIdx}\x_{\FreqIdx,\BlockIdx}\herm}, \vspace{-6pt}
			\label{eq:weighted_covMat}
		\end{equation}
		where \vspace{-6pt}
		\begin{equation}
			r_{\BlockIdx}^\ChannelIdx\left(\SetW_\ChannelIdx^{(l)}\right) = \left\Vert \yFreqVec_{\ChannelIdx,\BlockIdx}^{(l)} \right\Vert_2 = \sqrt{\sum_{\FreqIdx = 1}^{\FreqIdxMax} \left\vert \left(\w_\FreqIdx^{\ChannelIdx,(l)}\right)\herm\x_{\FreqIdx,\BlockIdx} \right\vert^2}.
			\label{eq:signal_energy}
		\end{equation}
		Using the inequality \eqref{eq:ono_inequality}, the following upper bound for the cost function \eqref{eq:proposed_cost_function} can be derived \Walter{by subtracting $\WASPAA{2}\sum_{\FreqIdx=1}^{\FreqIdxMax} \log \vert \det \W_\FreqIdx \vert$ and adding $J_\text{prior}(\SetW)$ on both sides of \eqref{eq:ono_inequality}} \vspace{-2pt}
		\WASPAA{\begin{align}
			&Q\left(\SetW\vert\SetW^{(l)}\right)  = \sum_{\FreqIdx=1}^{\FreqIdxMax}\Bigg[ -\WASPAA{2}\log \vert \det \W_\FreqIdx \vert + \sum_{\ChannelIdx=1}^{\ChannelIdxMax}\Bigg(\frac{R_\ChannelIdx\left(\SetW_\ChannelIdx^{(l)}\right)}{\FreqIdxMax} \dots \vspace{-3pt}\\
			&  \qquad + \left(\w_\FreqIdx^\ChannelIdx\right)\herm \left(\V_\FreqIdx^\ChannelIdx\left(\SetW_\ChannelIdx^{(l)}\right)+\frac{\lambda_\text{E}\mathbf{I}+\SteeringVector_\FreqIdx^{\WASPAA{\ChannelIdx}}(\SteeringVector_\FreqIdx^{\WASPAA{\ChannelIdx}})\herm}{\sigma^2_\FreqIdx}\right)\w_\FreqIdx^\ChannelIdx  \Bigg)\Bigg],\notag
		\end{align}}
		with $J(\SetW) = Q\left(\SetW\vert\SetW^{(l)}\right) $ iff $\SetW = \SetW^{(l)}$. \vspace{-2pt}
		\begin{algorithm}
			\caption{Informed \ac{IVA}}
			\label{alg:pseudocode}
			\begin{algorithmic}
				\STATE \textbf{INPUT:} $\X$, $L$, $\{\sigma^2_\FreqIdx\}_{\FreqIdx\in\SetK}$
				\STATE \textbf{Initalize:} $\W_\FreqIdx^{(0)} = \mathbf{I}_\ChannelIdxMax$ $\forall \FreqIdx \in \SetK$, $\y_{\FreqIdx,\BlockIdx} = \x_{\FreqIdx,\BlockIdx}$ $\forall \FreqIdx\in\SetK,\BlockIdx\in\SetN$
				\FOR{$l=1$ \TO $L$} 
				\FOR{$\ChannelIdx=1$ \TO $\ChannelIdxMax$} 
				\STATE Estimate energy of demixed signals by \eqref{eq:signal_energy}
				$\forall \BlockIdx\in\SetN$
				\FOR{$\FreqIdx=1$ \TO $\FreqIdxMax$}
				\STATE Estimate weighted covariance matrix \eqref{eq:weighted_covMat}
				\IF{\WASPAA{$\ChannelIdx \in \mathcal{I}$}} 
				\STATE{Update \WASPAA{constrained} demixing vector \eqref{eq:update_dirConstraint}, \eqref{eq:normilization_dirConstraint}}
				\ELSE
				\STATE{Update demixing vectors without constraint \eqref{eq:update}, \eqref{eq:normalization}} 
				\ENDIF	
				\ENDFOR
				\ENDFOR
				\ENDFOR
				\STATE \textbf{OUTPUT:} $\SetW$
			\end{algorithmic}
		\end{algorithm}
		\subsection{Minimization of the Upper Bound}
		\begin{figure*}
			\setlength\fwidth{0.25\textwidth}
			\begin{minipage}[t]{180pt}
				\begin{tikzpicture}

\begin{axis}[%
width=1\fwidth,
height=0.6\fwidth,
at={(0\fwidth,0\fwidth)},
scale only axis,
xmin=10,
xmax=30,
ymin=8,
ymax=30,
ylabel style={font=\color{white!15!black},yshift = -15pt},
ylabel={SIR/$[\mathrm{dB}] \rightarrow$},
title style = {yshift = -5pt},
title = {\textbf{Room 1:} $\mathbf{T_{60} = 50\,\mathbf{ms}}$},
axis background/.style={fill=white},
xmajorgrids,
ymajorgrids,
legend style={at={(0.03,0.8)},anchor=west,legend cell align=left, align=left, draw=white!15!black},
xtick = {10,20,30}, xticklabels = {,,}
]
\addplot [color=black, mark=x, mark size = 4pt, mark options={solid, black}, line width = 2pt]
  table[row sep=crcr]{%
10	15.4222398734279\\
20	18.5305648609945\\
30	29.225983156377\\
};

\addplot [color=black, dashed, mark=triangle, mark size = 4pt, mark options={solid, black}, line width = 2pt]
  table[row sep=crcr]{%
10	14.2851075025187\\
20	15.7393366331083\\
30	18.1526987828928\\
};

\addplot [color=black, dotted, mark=diamond, mark size = 4pt, mark options={solid, black}, line width = 2pt]
  table[row sep=crcr]{%
10	21.3456200002686\\
20	24.6222627361794\\
30	24.2623799847306\\
};

\end{axis}
\end{tikzpicture}%
\vspace{-8pt}
				\begin{tikzpicture}

\begin{axis}[%
width=1\fwidth,
height=0.6\fwidth,
at={(0\fwidth,0\fwidth)},
scale only axis,
xmin=10,
xmax=30,
xlabel style={font=\color{white!15!black},yshift = 5pt},
xlabel={SNR/$[\mathrm{dB}] \rightarrow$},
ymin=2,
ymax=21,
ylabel style={font=\color{white!15!black},yshift = -15pt},
ylabel={SDR/$[\mathrm{dB}] \rightarrow$},
axis background/.style={fill=white},
xmajorgrids,
ymajorgrids,
legend style={at={(0.03,0.8)},anchor=west,legend cell align=left, align=left, draw=white!15!black},
xtick = {10,20,30}
]
\addplot [color=black, mark=x, mark size = 4pt, mark options={solid, black}, line width = 2pt]
  table[row sep=crcr]{%
10	7.09459251911634\\
20	14.1835325039229\\
30	20.2236816256992\\
};

\addplot [color=black, dashed, mark=triangle, mark size = 4pt, mark options={solid, black}, line width = 2pt]
  table[row sep=crcr]{%
10	5.78136930252085\\
20	10.9373146464671\\
30	14.327029825572\\
};

\addplot [color=black, dotted, mark=diamond, mark size = 4pt, mark options={solid, black}, line width = 2pt]
  table[row sep=crcr]{%
10	5.92979398973671\\
20	12.700268426115\\
30	17.3911780388094\\
};

\end{axis}
\end{tikzpicture}%
			\end{minipage}
			\begin{minipage}[t]{170pt}
				\begin{tikzpicture}

\begin{axis}[%
width=1\fwidth,
height=0.6\fwidth,
at={(0\fwidth,0\fwidth)},
scale only axis,
xmin=10,
xmax=30,
ymin=8,
ymax=30,
title style = {yshift = -5pt},
title = {\textbf{Room 2:} $\mathbf{T_{60} = 200\,\mathbf{ms}}$},
axis background/.style={fill=white},
xmajorgrids,
ymajorgrids,
legend style={at={(0.03,0.8)},anchor=west,legend cell align=left, align=left, draw=white!15!black},
xtick = {10,20,30}, xticklabels = {,,}
]
\addplot [color=black, mark=x, mark size = 4pt, mark options={solid, black}, line width = 2pt]
  table[row sep=crcr]{%
10	14.4919951951686\\
20	18.6239925353541\\
30	18.3407532474383\\
};

\addplot [color=black, dashed, mark=triangle, mark size = 4pt, mark options={solid, black}, line width = 2pt]
  table[row sep=crcr]{%
10	8.45465534566638\\
20	10.7838566442755\\
30	12.4319602320364\\
};

\addplot [color=black, dotted, mark=diamond, mark size = 4pt, mark options={solid, black}, line width = 2pt]
  table[row sep=crcr]{%
10	17.4774684731697\\
20	16.6308053791462\\
30	16.4702260019411\\
};

\end{axis}
\end{tikzpicture}%
\vspace{-8pt}
				\begin{tikzpicture}

\begin{axis}[%
width=1\fwidth,
height=0.6\fwidth,
at={(0\fwidth,0\fwidth)},
scale only axis,
xmin=10,
xmax=30,
xlabel style={font=\color{white!15!black},yshift = 5pt},
xlabel={SNR/$[\mathrm{dB}] \rightarrow$},
ymin=2,
ymax=21,
axis background/.style={fill=white},
xmajorgrids,
ymajorgrids,
legend style={at={(0.03,0.8)},anchor=west,legend cell align=left, align=left, draw=white!15!black},
xtick = {10,20,30}
]
\addplot [color=black, mark=x, mark size = 4pt, mark options={solid, black}, line width = 2pt]
  table[row sep=crcr]{%
10	6.40214876506018\\
20	10.2761785635382\\
30	11.4557403654773\\
};

\addplot [color=black, dashed, mark=triangle, mark size = 4pt, mark options={solid, black}, line width = 2pt]
  table[row sep=crcr]{%
10	4.45168709672133\\
20	8.16741370755283\\
30	9.57540127151193\\
};

\addplot [color=black, dotted, mark=diamond, mark size = 4pt, mark options={solid, black}, line width = 2pt]
  table[row sep=crcr]{%
10	4.43178303056967\\
20	7.16262669880607\\
30	8.05418235684549\\
};

\end{axis}
\end{tikzpicture}%
			\end{minipage}
			\begin{minipage}[t]{170pt}
				\begin{tikzpicture}

\begin{axis}[%
width=1\fwidth,
height=0.6\fwidth,
at={(0\fwidth,0\fwidth)},
scale only axis,
xmin=10,
xmax=30,
ymin=8,
ymax=30,
title style = {yshift = -5pt},
title = {\textbf{Room 3:} $\mathbf{T_{60} = 400\,\mathbf{ms}}$},
axis background/.style={fill=white},
xmajorgrids,
ymajorgrids,
legend style={at={(0.03,0.8)},anchor=west,legend cell align=left, align=left, draw=white!15!black},
xtick = {10,20,30}, xticklabels = {,,}
]
\addplot [color=black, mark=x, mark size = 4pt, mark options={solid, black}, line width = 2pt]
  table[row sep=crcr]{%
10	13.0098933657116\\
20	17.1619313930244\\
30	16.538518090896\\
};

\addplot [color=black, dashed, mark=triangle, mark size = 4pt, mark options={solid, black}, line width = 2pt]
  table[row sep=crcr]{%
10	9.53336642284061\\
20	11.5139452856018\\
30	12.9851978189523\\
};

\addplot [color=black, dotted, mark=diamond, mark size = 4pt, mark options={solid, black}, line width = 2pt]
  table[row sep=crcr]{%
10	15.6464117361853\\
20	15.2825096454977\\
30	14.9110036272078\\
};

\end{axis}
\end{tikzpicture}%
\vspace{-8pt}
				\begin{tikzpicture}

\begin{axis}[%
width=1\fwidth,
height=0.6\fwidth,
at={(0\fwidth,0\fwidth)},
scale only axis,
xmin=10,
xmax=30,
xlabel style={font=\color{white!15!black},yshift = 5pt},
xlabel={SNR/$[\mathrm{dB}] \rightarrow$},
ymin=2,
ymax=21,
axis background/.style={fill=white},
xmajorgrids,
ymajorgrids,
legend style={at={(0.4,0.7)},anchor=west,legend cell align=left, align=left, draw=white!15!black},
xtick = {10,20,30}
]
\addplot [color=black, mark=x, mark size = 4pt, mark options={solid, black}, line width = 2pt]
  table[row sep=crcr]{%
10	4.00092429135117\\
20	5.93658587364447\\
30	6.13261609728267\\
};
\addlegendentry{auxIVA}

\addplot [color=black, dashed, mark=triangle, mark size = 4pt, mark options={solid, black}, line width = 2pt]
  table[row sep=crcr]{%
10	3.00940715446647\\
20	5.50097408103756\\
30	6.37999348928424\\
};
\addlegendentry{GC gradIVA}

\addplot [color=black, dotted, mark=diamond, mark size = 4pt, mark options={solid, black}, line width = 2pt]
  table[row sep=crcr]{%
10	2.93850732891987\\
20	4.44940760314243\\
30	4.75695959328216\\
};
\addlegendentry{GC auxIVA}

\end{axis}
\end{tikzpicture}%
			\end{minipage}\vspace{-8pt}
			\caption{\ac{SIR} values (first row) and \ac{SDR} values (second row) of the proposed algorithm GC auxIVA and the two benchmark algorithms auxIVA and GC gradIVA averaged over different directional priors and source \acp{DOA} for three different rooms.}\vspace{-8pt}
			\label{fig:results}
		\end{figure*}
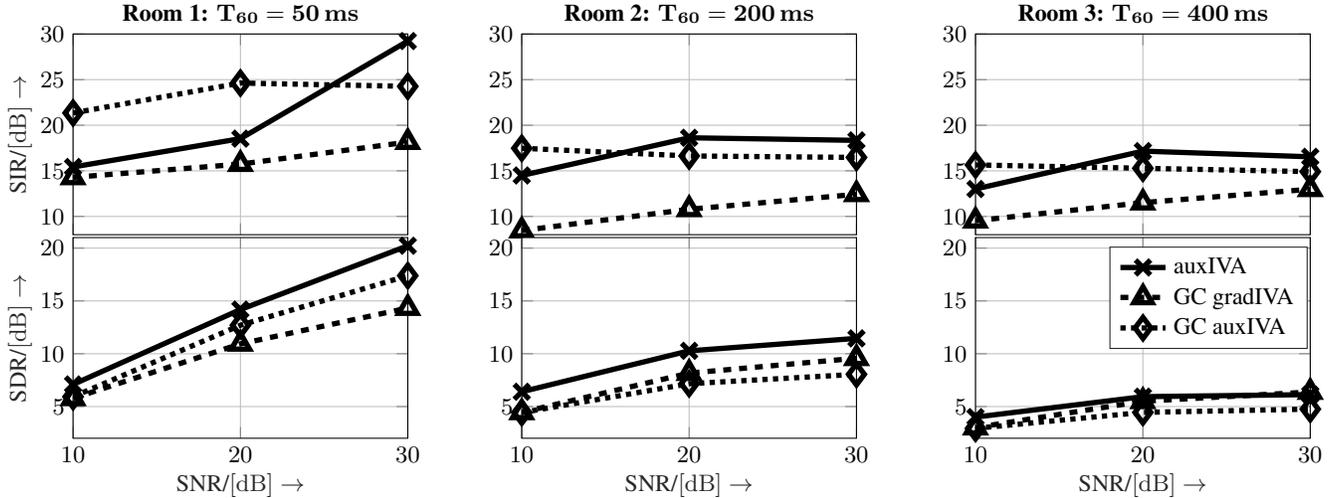
		In order to construct update rules following the \ac{MM} philosophy, we minimize the upper bound, which 
yields the following conditions for the demixing matrices, \WASPAA{where $q \in \{1,\dots,\ChannelIdxMax\}$} \vspace{-2pt}
		\begin{align}
			&\left(\w_\FreqIdx^q\right)\herm \V_\FreqIdx^\ChannelIdx\left(\SetW_\ChannelIdx^{(l)}\right)\w_\FreqIdx^\ChannelIdx \overset{!}{=}\delta_{\ChannelIdx q} \qquad\qquad\qquad\qquad \text{for} \  \WASPAA{\ChannelIdx\notin \mathcal{I}}\nonumber\\
			&\left(\w_\FreqIdx^q\right)\herm \left[\V_\FreqIdx^\ChannelIdx\left(\SetW_\ChannelIdx^{(l)}\right) + \WASPAA{\frac{\lambda_\text{E}\mathbf{I}+\SteeringVector_\FreqIdx^{\WASPAA{\ChannelIdx}}(\SteeringVector_\FreqIdx^{\WASPAA{\ChannelIdx}})\herm}{\sigma^2_\FreqIdx}}\right]\w_\FreqIdx^\ChannelIdx \overset{!}{=}\delta_{\ChannelIdx q} \quad\text{for} \ \WASPAA{\ChannelIdx \in \mathcal{I}},\nonumber \vspace{-6pt}
		\end{align}
		where $\delta_{\ChannelIdx q}$ denotes the Kronecker delta, i.e., $\delta_{\ChannelIdx q} = 1$ iff $\ChannelIdx=q$ and $\delta_{\ChannelIdx q} = 0$ else. For \WalterTwo{solving} this problem, we adopt a sequential update strategy \cite{ono_stable_2011}, which results in the following update rules \WASPAA{for $k\in \mathcal{I}$} \vspace{-2pt}
		\begin{equation}
			\tilde{\w}_\FreqIdx^{\ChannelIdx,(l+1)} = \left(\W_\FreqIdx^{(l)}\left[\V_\FreqIdx^{\ChannelIdx,(l)}\left(\SetW_\ChannelIdx^{(l)}\right) + \WASPAA{\frac{\lambda_\text{E}\mathbf{I}+\SteeringVector_\FreqIdx^{\WASPAA{\ChannelIdx}}(\SteeringVector_\FreqIdx^{\WASPAA{\ChannelIdx}})\herm}{\sigma^2_\FreqIdx}}\right]\right)^{-1}\mathbf{e}_\ChannelIdx,
			\label{eq:update_dirConstraint}
		\end{equation} \vspace{-10pt}
		\begin{equation}
			\w_\FreqIdx^{\ChannelIdx,(l+1)} = \frac{\tilde{\w}_\FreqIdx^{\ChannelIdx,(l+1)}}{\sqrt{\left(\tilde{\w}_\FreqIdx^{\ChannelIdx,(l)}\right)\herm \left[\V_\FreqIdx^{\ChannelIdx,(l)}\left(\SetW_\ChannelIdx^{(l)}\right) + \WASPAA{\frac{\lambda_\text{E}\mathbf{I}+\SteeringVector_\FreqIdx^{\WASPAA{\ChannelIdx}}(\SteeringVector_\FreqIdx^{\WASPAA{\ChannelIdx}})\herm}{\sigma^2_\FreqIdx}}\right]\tilde{\w}_\FreqIdx^{\ChannelIdx,(l)}}}
			\label{eq:normilization_dirConstraint}
		\end{equation}
		and for \WASPAA{$\ChannelIdx\notin \mathcal{I}$} \vspace{-2pt}
		\begin{equation}
			\tilde{\w}_\FreqIdx^{\ChannelIdx,(l+1)} = \left(\W_\FreqIdx^{\ChannelIdx,(l)}\V_\FreqIdx^{\ChannelIdx,(l)}\left(\SetW_\ChannelIdx^{(l)}\right)\right)^{-1}\mathbf{e}_\ChannelIdx,
			\label{eq:update}
		\end{equation} \vspace{-9pt}
		\begin{equation}
			\w_\FreqIdx^{\ChannelIdx,(l+1)} = \frac{\tilde{\w}_\FreqIdx^{\ChannelIdx,(l+1)}}{\sqrt{\left(\tilde{\w}_\FreqIdx^{\ChannelIdx,(l)}\right)\herm \V_\FreqIdx^{\ChannelIdx,(l)}\left(\SetW_\ChannelIdx^{(l)}\right)\tilde{\w}_\FreqIdx^{\ChannelIdx,(l)}}},
			\label{eq:normalization}
		\end{equation}
		where $\mathbf{e}_\ChannelIdx$ denotes the canonical unit vector with a one as its $\ChannelIdx$th entry.	
		Algorithm~\ref{alg:pseudocode} summarizes the proposed method for estimating the demixing matrices $\SetW$. The final step is the demixing of the recorded signals according to \eqref{eq:demixing_equation}.
		\section{\Walter{Experiments}}
		\label{sec:experiments}
		To show the efficacy of the proposed algorithm, we carried out experiments based on measured \acp{RIR} of three rooms, i.e., a low-reverberant chamber (Room 1, $T_{60} = 50\,\mathrm{ms}$) and two meeting rooms (Room 2 \& 3, $T_{60} = 200\,\mathrm{ms}, 400\,\mathrm{ms}$) with a microphone pair of $0.21\,\mathrm{m}$ spacing. Clean speech signals of a female and a male speaker \WalterTwo{$(\ChannelIdxMax = 2)$} are convolved with \Walter{the according \acp{RIR}, mixed, and} distorted by additive white Gaussian noise to simulate the microphone signals. These are transformed into the \ac{STFT} domain by employing a Hamming window of length $2048$ and $50\%$ overlap at a sampling \WalterTwo{rate} of $16\,\mathrm{kHz}$. 
		
		In the following, we compare the performance of the proposed algorithm (GC auxIVA)\WASPAA{, with a prior on the first source and an uninformative prior on the second source,} with \ac{auxIVA} \cite{ono_stable_2011} and the \WalterTwo{\ac{GC}} gradient-based \ac{IVA} algorithm \cite{vincent_geometrically_2015} (GC gradIVA), which steers a spatial \WalterTwo{one} into the target direction. All methods use the source model $G\left(r_{\BlockIdx}^\ChannelIdx\right) = r_{\BlockIdx}^\ChannelIdx$ and are evaluated with a sufficiently large number of iterations to ensure convergence ($L=100$ for auxIVA and GC auxIVA and $L=350$ for GC gradIVA). The variance of the Gaussian prior \eqref{eq:directional_prior} has been chosen to be constant for all frequencies $\sigma_\FreqIdx^2 = \sigma^2 = 40$ \WASPAA{and the filter energy penalty parameter is chosen to be $\lambda_\text{E} = 10^{-3}$}. The stepsize for GC gradIVA has been set to $0.05$ and the weighting of the directional constraint to $0.5$. These values yielded the fastest convergence and the smallest influence of the regularizing term while still resolving the outer permutation problem.
		
		To quantify the performance of the proposed algorithm, we chose \acp{RIR} measured at $1\,\mathrm{m}$ distance from the microphone array and $45^\circ/135^\circ$, $45^\circ/90^\circ$ and $20^\circ/160^\circ$ for Room 1 and $50^\circ/130^\circ$, $50^\circ/90^\circ$ and $10^\circ/170^\circ$ for Room 2 \& 3 w.r.t. the array axis \Walter{corresponding to averaged \acp{DRR} of $6.8\,\mathrm{dB}$, $4.5\,\mathrm{dB}$ and $3.3\,\mathrm{dB}$, respectively}. \Walter{We} synthesized microphone signals corresponding to \ac{SNR} values of $10\,\mathrm{dB}, 20\,\mathrm{dB}, 30\,\mathrm{dB}$. For each configuration, solutions with a constraint on each source direction are computed and the resulting \ac{SIR} and \ac{SDR} values computed by employing the toolbox \cite{vincent_performance_2006} are averaged over all source configurations and directional constraints to yield the results for auxIVA, GC auxIVA and GC gradIVA depicted in Fig.~\ref{fig:results}. \Walter{Note that the outer permutation problem is solved by GC auxIVA and GC gradIVA algorithmically, whereas it is not solved by \ac{auxIVA} \WalterTwo{which is the main motivation for considering constrained \ac{IVA} algorithms}.} It can be seen that the proposed GC auxIVA obtains a higher \ac{SIR} than GC gradIVA in all scenarios and is comparable with auxIVA. The \ac{SDR} of GC auxIVA is slightly lower than auxIVA \Walter{due to the free-field prior}\WalterTwo{,} but comparable with GC gradIVA in general. \Walter{The \ac{SDR} is decreasing for all algorithms for increasing $T_{60}$ due to the decreasing \ac{DRR}. Note that for disambiguating \WASPAA{$\ChannelIdxMax$ sources $\vert\mathcal{I}\vert = \ChannelIdxMax-1$} prior terms of the form \eqref{eq:directional_prior} can be used.}
		
\begin{figure}\vspace{-5pt}
			\setlength\fwidth{0.4\textwidth}
			\input{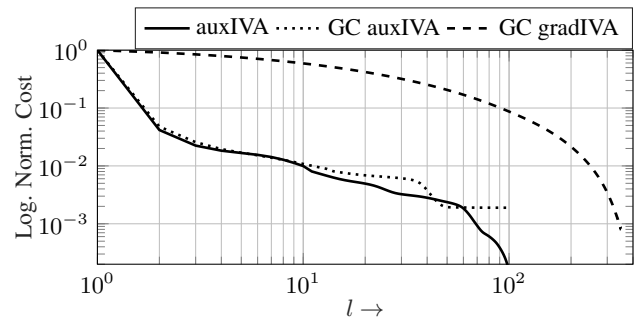}\vspace{-10pt}
			\caption{\Walter{Exemplary} \WalterTwo{behaviour} of the \Walter{logarithmic} normalized \Walter{\ac{IVA}} cost function values \WASPAA{(without considering the prior)} for the three investigated algorithms. \vspace{-10pt}}
			\label{fig:costs}
		\end{figure}

		Finally, \WalterTwo{the convergence speed of the investigated algorithms is compared.} To this end, we show a typical curve of the cost function values \WASPAA{normalized to the inital cost} over the iterations $l$ in a semi-logarithmic scale in Fig.~\ref{fig:costs}. It can be seen that auxIVA and GC auxIVA \Walter{exhibit almost identical and} a much faster convergence speed than GC gradIVA. One iteration of auxIVA or GC auxIVA needs on average $0.24\,\mathrm{s}$ and GC gradIVA $0.15\,\mathrm{s}$ on a notebook with an Intel Core i7-5600U CPU. Hence, one iteration of GC auxIVA is computationally slightly more demanding, however, needs much less iterations to converge than GC gradIVA and is hence computationally cheaper.
		\section{Conclusion}
		\label{sec:conclusion}
		We presented a \ac{MAP} derivation of \ac{IVA} \Walter{including} a \Walter{directional} prior over the \Walter{demixing filters} \Walter{to solve the outer permutation problem of \ac{BSS} algorithms}. The resulting cost function is efficiently solved by an \ac{MM} algorithm achieving \Walter{dramatically} faster convergence speed and higher interference suppression than \Walter{a comparable} state-of-the-art competing method.
		Future work may include the discussion of \Walter{other priors} for the source direction including priors which steer a spatial one into the direction of interest. Additionally, the derivation of a \WalterTwo{fully Bayesian} approach including hyperpriors over the source\WalterTwo{, e.g., its variance,} may be one of the next steps.
\bibliographystyle{IEEEbib}
\bibliography{literature}

\end{document}